\documentclass{article}
\usepackage{amsfonts}
\usepackage{amsmath}
\usepackage{amssymb}

\newtheorem{theorem}{Theorem}

\newtheorem{remark}[theorem]{Remark}

\input{tcilatex}
\begin{document}

\title{On Integrating the Left-Flat Vacuum Einstein Equations}
\author{Ezra T. Newman \\
Dept of Physics and Astronomy, University of Pittsburgh\\
Pittsburgh, PA, 15260, USA}

\begin{abstract}
Considering the spin-coefficient version of the left-flat vacuum Einstein
equations, all but one of the fifty equations can be explicitly integrated
via the introduction of five spin-weight s=-2 complex potentials. \ The
final equation is a non-linear wave equation for the last of the potentials.
Solutions to this equation determine solutions for the entire system.

Solutions for several special cases are obtained.
\end{abstract}

\date{March 7, 201}

\section{\textbf{I. Introduction}}

\subsection{ Background}

The self-dual/left flat (or anti-self dual/right/flat) Ricci-flat equations
and their spaces with associated metrics has been a subject of considerable
interest since the late 1970s. The very large literature has been associated
with a variety of linked interests.

They play a fundamental role in the properties of real asymptotically flat
space-times via solutions to the `good-cut' equation$^{1,2,3,4,5,6,7}$, they
define Penrose's asymptotic twistor space$^{8,9,10}$, they (the solutions)
are intimately related to (in fact are) Penrose's non-linear graviton
fields, the equations themselves are studied as a beautiful example of a
non-linear integrable system$^{11,12,13}$ and finally the Euclidean
versions, the so-called gravitational instantons$^{14,15}\ $(the general
relativistic generalization of the Yang-Mills instantons) have played a role
in attempts at understanding quantum gravity.

From the beginning there have been a variety of attempts at integrating the
field equations and even now - 40 years after their popularity began - only
a relatively small number of solutions appear to be known. The first
solutions - a very limited number - were obtained via the 'good cut'
equation. The best know is the Sparling-Tod metric$^{6}$/ Eguchi-Hansen$%
^{7}\ $metric. Many others were discovered by imposing special conditions as
for example algebraic specialness$^{16}$ or (Killing) symmetries$^{17}$. \
In addition there is a large literature on self-dual gravitational instantons%
$^{14}.$

To our knowledge there have been very few attacks on the problem of finding
(or even of studying)\ the general solution of the self-dual Einstein
equations on general complex four dimensional manifolds - either globally or
locally$^{4}$. \ It is the purpose of this note to return to this issue.

Although all H-spaces (i.e., spaces arising from solutions of the good-cut
equation) are self-dual spaces, it is not yet clear whether all self-dual
spaces are H-spaces. Nevertheless we will not make a distinction here
between the two.

\subsection{Modus Operandi}

We begin, in Sec.II, with the fully general Einstein equations in the
spin-coefficient formalism written as complex equations on a thickened
region of $C^{4}\ $in a neighborhood of the real 'slice' $R^{4}.\ $In the
standard "real" version of the formalism, the basic field variables, i.e.,
the spin-coefficients, the metric variables and Weyl tensor components, are,
in general, complex. In the field equations themselves, their complex
conjugates appear symmetrically so that the equations themselves, as a
complete set are real. In our present (SD) version however, the previous
"complex conjugates" are now completely freed up from their conjugate
counterparts and are independent variables. These field equations are then
reduced by setting the anti-self-dual part of the Weyl tensor to zero,
thereby imposing the self-dual condition on the equations.

A coordinate system is introduced, in Sec.II, by choosing a world-line, $%
\mathfrak{L,}$ in the thickened region of C$^{4}:$ with the (complex) $u$,
being the `complex time' along the world line. The null cones with apex on
the line are labeled by $u.\ $The complex sphere of null directions at the
apex of each cone (and their associated null geodesics) are labeled by the
complex stereographic coordinates ($\zeta ,\widetilde{\zeta }$) while the
affine parameter along the null geodesics is given by $r$. Note that both $%
u\ $and$\ r$ have values close to the real while $\widetilde{\zeta }\ $is
close to the complex conjugate $\overline{\zeta }.$

From these imposed conditions we easily show, Sec.II, that the complex
divergence, $\rho ,$ of the null geodesics with apex on $\mathfrak{L}$ is
given by $\rho =-r^{-1}\ $and the left shear by $\sigma =0.$

\noindent \qquad At this point the entire set of field equations (about 50)
are divided into three sets: (1) the radial equations which involve the $r\ $%
derivative, (2) those involving the angular derivatives, i.e., ($\zeta ,%
\widetilde{\zeta }$) and (3) those that contain the time-derivative, $u\ \ $%
-\ \ about(14) of them.

\subsection{Results}

\qquad What is a bit surprising is that every one of the $r$-derivative
equations can be exactly integrated in terms of four spin-weight ($s=-2)$,
potentials, ($\gamma _{0},\gamma _{1},\gamma _{2},\gamma _{3},\gamma _{4}$)$%
,\ $where each can be expressed as $r$-derivatives of the next one, so that
the only independent one is $\gamma _{4}.$

\qquad The angular equations then are then used to establish relationships
between some of the free "constants" of the $r$-integrations.

\qquad Finally when all these results are inserted into the (14)
time-derivative equations we discover that they are all closely related and
can be reduced to a \textit{single complex spin-weight (}$s=-2)$\textit{,
non-linear wave equation}\ for the last potential $\gamma _{4}.\ $This last
equation carries all the evolution information for the entire set of
equations.

Though we do not know of a way to give general solutions to this equation,
it is quite easy to produce many special solutions.

\section{\noindent \textbf{II. \ The Field Equations}}

\subsection{Spin-Coefficient version of Einstein equations}

The familiar spin-coefficient version of the Einstein equations$^{18}$- with
their "conjugates" explicitly inserted and written out in detail - are:

The metric equations:

\begin{eqnarray}
\Delta l^{a}-Dn^{a} &=&(\gamma +\overline{\gamma })l^{a}+(\epsilon +%
\overline{\epsilon })n^{a}-(\tau +\overline{\pi })\overline{m}^{a}-(%
\overline{\tau }+\pi )m^{a},  \label{metric eqs} \\
\delta l^{a}-Dm^{a} &=&(\overline{\alpha }+\beta -\overline{\pi }%
)l^{a}+\kappa n^{a}-\sigma \overline{m}^{a}-(\overline{\rho }+\epsilon -%
\overline{\epsilon })m^{a},  \notag \\
\overline{\delta }l^{a}-D\overline{m}^{a} &=&(\alpha +\overline{\beta }-\pi
)l^{a}+\overline{\kappa }n^{a}-\overline{\sigma }m^{a}-(\rho +\overline{%
\epsilon }-\epsilon )\overline{m}^{a},  \notag \\
\delta n^{a}-\Delta m^{a} &=&-\nu l^{a}+(\tau -\overline{\alpha }-\beta
)n^{a}+\overline{\lambda }\overline{m}^{a}+(\mu -\gamma +\overline{\gamma }%
)m^{a},  \notag \\
\overline{\delta }n^{a}-\Delta \overline{m}^{a} &=&-\overline{\nu }l^{a}+(%
\overline{\tau }-\alpha -\overline{\beta })n^{a}+\lambda m^{a}+(\overline{%
\mu }-\overline{\gamma }+\gamma )\overline{m}^{a}  \notag \\
\overline{\delta }m^{a}-\delta \overline{m}^{a} &=&(\overline{\mu }-\mu
)l^{a}+(\overline{\rho }-\rho )n^{a}-(\overline{\alpha }-\beta )\overline{m}%
^{a}+(\alpha -\overline{\beta })m^{a}.  \notag
\end{eqnarray}

The spin-coefficient equations:

\begin{eqnarray}
\Delta \lambda -\overline{\delta }\nu &=&-(\mu +\overline{\mu }+3\gamma -%
\overline{\gamma })\lambda +(3\alpha +\overline{\beta }+\pi -\overline{\tau }%
)\nu -\Psi _{4}  \label{sp.coeff} \\
\Delta \overline{\lambda }-\delta \overline{\nu } &=&-(\mu +\overline{\mu }+3%
\overline{\gamma }-\gamma )\overline{\lambda }+(3\overline{\alpha }+\beta +%
\overline{\pi }-\tau )\overline{\nu }-\overline{\Psi }_{4}  \notag \\
\delta \rho -\overline{\delta }\sigma &=&\rho (\overline{\alpha }+\beta
)-\sigma (3\alpha -\overline{\beta })+(\rho -\overline{\rho })\tau +(\mu -%
\overline{\mu })\kappa -\Psi _{1}  \notag \\
\overline{\delta }\overline{\rho }-\delta \overline{\sigma } &=&\overline{%
\rho }(\alpha +\overline{\beta })-\overline{\sigma }(3\overline{\alpha }%
-\beta )+(\overline{\rho }-\rho )\overline{\tau }+(\overline{\mu }-\mu )%
\overline{\kappa }-\overline{\Psi }_{1}  \notag \\
\delta \alpha -\overline{\delta }\beta &=&\mu \rho -\lambda \sigma +\alpha 
\overline{\alpha }+\beta \overline{\beta }-2\alpha \beta +\gamma (\rho -%
\overline{\rho })+\epsilon (\mu -\overline{\mu })-\Psi _{2}  \notag \\
\overline{\delta }\overline{\alpha }-\delta \overline{\beta } &=&\overline{%
\mu }\overline{\rho }-\overline{\lambda }\overline{\sigma }+\alpha \overline{%
\alpha }+\beta \overline{\beta }-2\overline{\alpha }\overline{\beta }-%
\overline{\gamma }(\rho -\overline{\rho })-\overline{\epsilon }(\mu -%
\overline{\mu })-\overline{\Psi }_{2}  \notag \\
\delta \lambda -\overline{\delta }\mu &=&(\rho -\overline{\rho })\nu +(\mu -%
\overline{\mu })\pi +\mu (\alpha +\overline{\beta })+\lambda (\overline{%
\alpha }-3\beta )-\Psi _{3}  \notag \\
\overline{\delta }\overline{\lambda }-\delta \overline{\mu } &=&-(\rho -%
\overline{\rho })\overline{\nu }-(\mu -\overline{\mu })\overline{\pi }+%
\overline{\mu }(\overline{\alpha }+\beta )+\overline{\lambda }(\alpha -3%
\overline{\beta })-\overline{\Psi }_{3}  \notag \\
\delta \nu -\Delta \mu &=&\mu ^{2}+\lambda \overline{\lambda }+\mu (\gamma +%
\overline{\gamma })-\overline{\nu }\pi +\nu (\tau -3\beta -\overline{\alpha }%
)  \notag \\
\overline{\delta }\overline{\nu }-\Delta \overline{\mu } &=&\overline{\mu }%
^{2}+\lambda \overline{\lambda }+\overline{\mu }(\gamma +\overline{\gamma }%
)-\nu \overline{\pi }+\overline{\nu }(\overline{\tau }-3\overline{\beta }%
-\alpha )  \notag \\
\delta \gamma -\Delta \beta &=&\gamma (\tau -\overline{\alpha }-\beta )+\mu
\tau -\sigma \nu -\epsilon \overline{\nu }-\beta (\gamma -\overline{\gamma }%
-\mu )+\alpha \overline{\lambda }  \notag \\
\overline{\delta }\overline{\gamma }-\Delta \overline{\beta } &=&\overline{%
\gamma }(\overline{\tau }-\alpha -\overline{\beta })+\overline{\mu }%
\overline{\tau }-\overline{\sigma }\overline{\nu }-\overline{\epsilon }\nu -%
\overline{\beta }(\overline{\gamma }-\gamma -\overline{\mu })+\overline{%
\alpha }\lambda  \notag \\
\delta \tau -\Delta \sigma &=&\mu \sigma +\rho \overline{\lambda }+\tau
(\tau +\beta -\overline{\alpha })-\sigma (3\gamma -\overline{\gamma }%
)-\kappa \overline{\nu }  \notag \\
\overline{\delta }\overline{\tau }-\Delta \overline{\sigma } &=&\overline{%
\mu }\overline{\sigma }+\overline{\rho }\lambda +\overline{\tau }(\overline{%
\tau }+\overline{\beta }-\alpha )-\overline{\sigma }(3\overline{\gamma }%
-\gamma )-\overline{\kappa }\nu  \notag \\
\Delta \rho -\overline{\delta }\tau &=&-\rho \overline{\mu }-\sigma \lambda
+\tau (\overline{\beta }-\alpha -\overline{\tau })+(\gamma +\overline{\gamma 
})\rho +\kappa \nu -\Psi _{2}  \notag \\
\Delta \overline{\rho }-\delta \overline{\tau } &=&-\overline{\rho }\mu -%
\overline{\sigma }\overline{\lambda }+\overline{\tau }(\beta -\overline{%
\alpha }-\tau )+(\gamma +\overline{\gamma })\overline{\rho }+\overline{%
\kappa }\overline{\nu }-\overline{\Psi }_{2}  \notag \\
\Delta \alpha -\overline{\delta }\gamma &=&\nu (\rho +\epsilon )-\lambda
(\tau +\beta )+\alpha (\overline{\gamma }-\overline{\mu })+\gamma (\overline{%
\beta }-\overline{\tau })-\Psi _{3}  \notag \\
\Delta \overline{\alpha }-\delta \overline{\gamma } &=&\overline{\nu }(%
\overline{\rho }+\overline{\epsilon })-\overline{\lambda }(\overline{\tau }+%
\overline{\beta })+\overline{\alpha }(\gamma -\mu )+\overline{\gamma }(\beta
-\tau )-\overline{\Psi }_{3}  \notag
\end{eqnarray}%
\noindent

\begin{eqnarray}
D\rho -\overline{\delta }\kappa &=&\rho ^{2}+\sigma \overline{\sigma }%
+(\epsilon +\overline{\epsilon })\rho -\overline{\kappa }\tau -\kappa
(3\alpha +\overline{\beta }-\pi )  \notag \\
D\overline{\rho }-\delta \overline{\kappa } &=&\overline{\rho }^{2}+\sigma 
\overline{\sigma }+(\epsilon +\overline{\epsilon })\overline{\rho }-\kappa 
\overline{\tau }-\overline{\kappa }(3\overline{\alpha }+\beta -\overline{\pi 
})  \notag \\
D\sigma -\delta \kappa &=&(\rho +\overline{\rho })\sigma +(3\epsilon -%
\overline{\epsilon })\sigma -(\tau -\overline{\pi }+\overline{\alpha }%
+3\beta )\kappa +\Psi _{0}  \notag \\
D\overline{\sigma }-\overline{\delta }\overline{\kappa } &=&(\rho +\overline{%
\rho })\overline{\sigma }+(3\overline{\epsilon }-\epsilon )\overline{\sigma }%
-(\overline{\tau }-\pi +\alpha +3\overline{\beta })\overline{\kappa }+%
\overline{\Psi }_{0} \\
D\tau -\Delta \kappa &=&(\tau +\overline{\pi })\rho +(\overline{\tau }+\pi
)\sigma +(\epsilon -\overline{\epsilon })\tau -(3\gamma +\overline{\gamma }%
)\kappa +\Psi _{1}  \notag \\
D\overline{\tau }-\Delta \overline{\kappa } &=&(\overline{\tau }+\pi )%
\overline{\rho }+(\tau +\overline{\pi })\overline{\sigma }-(\epsilon -%
\overline{\epsilon })\overline{\tau }-(3\overline{\gamma }+\gamma )\overline{%
\kappa }+\overline{\Psi }_{1} \\
D\alpha -\overline{\delta }\epsilon &=&(\rho +\overline{\epsilon }-2\epsilon
)\alpha +\beta \overline{\sigma }-\overline{\beta }\epsilon -\kappa \lambda -%
\overline{\kappa }\gamma +(\epsilon +\rho )\pi  \notag \\
D\overline{\alpha }-\delta \overline{\epsilon } &=&(\overline{\rho }%
+\epsilon -2\overline{\epsilon })\overline{\alpha }+\overline{\beta }\sigma
-\beta \overline{\epsilon }-\overline{\kappa }\overline{\lambda }-\kappa 
\overline{\gamma }+(\overline{\epsilon }+\overline{\rho })\overline{\pi } \\
D\beta -\delta \epsilon &=&(\alpha +\pi )\sigma +(\overline{\rho }-\overline{%
\epsilon })\beta -(\mu +\gamma )\kappa -(\overline{\alpha }-\overline{\pi }%
)\epsilon +\Psi _{1}  \notag \\
D\overline{\beta }-\overline{\delta }\overline{\epsilon } &=&(\overline{%
\alpha }+\overline{\pi })\overline{\sigma }+(\rho -\epsilon )\overline{\beta 
}-(\overline{\mu }+\overline{\gamma })\overline{\kappa }-(\alpha -\pi )%
\overline{\epsilon }+\overline{\Psi }_{1} \\
D\gamma -\Delta \epsilon &=&(\tau +\overline{\pi })\alpha +(\overline{\tau }%
+\pi )\beta -(\epsilon +\overline{\epsilon })\gamma -(\gamma +\overline{%
\gamma })\epsilon +\tau \pi -\nu \kappa +\Psi _{2}  \notag \\
D\overline{\gamma }-\Delta \overline{\epsilon } &=&(\overline{\tau }+\pi )%
\overline{\alpha }+(\tau +\overline{\pi })\overline{\beta }-(\epsilon +%
\overline{\epsilon })\overline{\gamma }-(\gamma +\overline{\gamma })%
\overline{\epsilon }+\overline{\tau }\overline{\pi }-\overline{\nu }%
\overline{\kappa }+\overline{\Psi }_{2} \\
D\lambda -\overline{\delta }\pi &=&\rho \lambda +\overline{\sigma }\mu +\pi
^{2}+(\alpha -\overline{\beta })\pi -\nu \overline{\kappa }-(3\epsilon -%
\overline{\epsilon })\lambda  \notag \\
D\overline{\lambda }-\delta \overline{\pi } &=&\overline{\rho }\overline{%
\lambda }+\sigma \overline{\mu }+\overline{\pi }^{2}+(\overline{\alpha }%
-\beta )\overline{\pi }-\overline{\nu }\kappa -(3\overline{\epsilon }%
-\epsilon )\overline{\lambda } \\
D\mu -\delta \pi &=&\overline{\rho }\mu +\sigma \lambda +\pi \overline{\pi }%
-(\epsilon +\overline{\epsilon })\mu -\pi (\overline{\alpha }-\beta )-\nu
\kappa +\Psi _{2}  \notag \\
D\overline{\mu }-\overline{\delta }\overline{\pi } &=&\rho \overline{\mu }+%
\overline{\sigma }\overline{\lambda }+\pi \overline{\pi }-(\epsilon +%
\overline{\epsilon })\overline{\mu }-\overline{\pi }(\alpha -\overline{\beta 
})-\overline{\nu }\overline{\kappa }+\overline{\Psi }_{2} \\
D\nu -\Delta \pi &=&(\overline{\tau }+\pi )\mu +(\tau +\overline{\pi }%
)\lambda +(\gamma -\overline{\gamma })\pi -(3\epsilon +\overline{\epsilon }%
)\nu +\Psi _{3}  \notag \\
D\overline{\nu }-\Delta \overline{\pi } &=&(\tau +\overline{\pi })\overline{%
\mu }+(\overline{\tau }+\pi )\overline{\lambda }-(\gamma -\overline{\gamma })%
\overline{\pi }-(3\overline{\epsilon }+\epsilon )\overline{\nu }+\overline{%
\Psi }_{3}
\end{eqnarray}

and finally the Bianchi Identities:

\begin{eqnarray}
\overline{\delta }\Psi _{0}-D\Psi _{1} &=&(4\alpha -\pi )\Psi _{0}-2(2\rho
+\epsilon )\Psi _{1}+3\kappa \Psi _{2},  \label{BIA} \\
\overline{\delta }\Psi _{1}-D\Psi _{2} &=&\lambda \Psi _{0}+2(\alpha -\pi
)\Psi _{1}-3\rho \Psi _{2}+2\kappa \Psi _{3},  \notag \\
\overline{\delta }\Psi _{2}-D\Psi _{3} &=&2\lambda \Psi _{1}-3\pi \Psi
_{2}+2(\epsilon -\rho )\Psi _{3}+\kappa \Psi _{4},  \notag \\
\overline{\delta }\Psi _{3}-D\Psi _{4} &=&3\lambda \Psi _{2}-2(\alpha +2\pi
)\Psi _{3}+(4\epsilon -\rho )\Psi _{4},  \notag
\end{eqnarray}

\begin{eqnarray}
\delta \overline{\Psi }_{0}-D\overline{\Psi }_{1} &=&(4\overline{\alpha }-%
\overline{\pi })\overline{\Psi }_{0}-2(2\overline{\rho }+\overline{\epsilon }%
)\overline{\Psi }_{1}+3\overline{\kappa }\overline{\Psi }_{2},  \label{BIB}
\\
\delta \overline{\Psi }_{1}-D\overline{\Psi }_{2} &=&\overline{\lambda }%
\overline{\Psi }_{0}+2(\overline{\alpha }-\overline{\pi })\overline{\Psi }%
_{1}-3\overline{\rho }\overline{\Psi }_{2}+2\overline{\kappa }\overline{\Psi 
}_{3},  \notag \\
\delta \overline{\Psi }_{2}-D\overline{\Psi }_{3} &=&2\overline{\lambda }%
\overline{\Psi }_{1}-3\overline{\pi }\overline{\Psi }_{2}+2(\overline{%
\epsilon }-\overline{\rho })\overline{\Psi }_{3}+\overline{\kappa }\overline{%
\Psi }_{4},  \notag \\
\delta \overline{\Psi }_{3}-D\overline{\Psi }_{4} &=&3\overline{\lambda }%
\overline{\Psi }_{2}-2(\overline{\alpha }+2\overline{\pi })\overline{\Psi }%
_{3}+(4\overline{\epsilon }-\overline{\rho })\overline{\Psi }_{4},  \notag
\end{eqnarray}

\begin{eqnarray}
\Delta \Psi _{0}-\delta \Psi _{1} &=&(4\gamma -\mu )\Psi _{0}-2(2\tau +\beta
)\Psi _{1}+3\sigma \Psi _{2},  \label{BIC} \\
\Delta \Psi _{1}-\delta \Psi _{2} &=&\nu \Psi _{0}+2(\gamma -\mu )\Psi
_{1}-3\tau \Psi _{2}+2\sigma \Psi _{3},  \notag \\
\Delta \Psi _{2}-\delta \Psi _{3} &=&2\nu \Psi _{1}-3\mu \Psi _{2}+2(\beta
-\tau )\Psi _{3}+\sigma \Psi _{4},  \notag \\
\Delta \Psi _{3}-\delta \Psi _{4} &=&3\nu \Psi _{2}-2(\gamma +2\mu )\Psi
_{3}+(4\beta -\tau )\Psi _{4}.  \notag
\end{eqnarray}

\begin{eqnarray}
\Delta \overline{\Psi }_{0}-\overline{\delta }\overline{\Psi }_{1} &=&(4%
\overline{\gamma }-\overline{\mu })\overline{\Psi }_{0}-2(2\overline{\tau }+%
\overline{\beta })\overline{\Psi }_{1}+3\overline{\sigma }\overline{\Psi }%
_{2},  \label{BID} \\
\Delta \overline{\Psi }_{1}-\overline{\delta }\overline{\Psi }_{2} &=&%
\overline{\nu }\overline{\Psi }_{0}+2(\overline{\gamma }-\overline{\mu })%
\overline{\Psi }_{1}-3\overline{\tau }\overline{\Psi }_{2}+2\overline{\sigma 
}\overline{\Psi }_{3},  \notag \\
\Delta \overline{\Psi }_{2}-\overline{\delta }\overline{\Psi }_{3} &=&2%
\overline{\nu }\overline{\Psi }_{1}-3\overline{\mu }\overline{\Psi }_{2}+2(%
\overline{\beta }-\overline{\tau })\overline{\Psi }_{3}+\overline{\sigma }%
\overline{\Psi }_{4},  \notag \\
\Delta \overline{\Psi }_{3}-\overline{\delta }\overline{\Psi }_{4} &=&3%
\overline{\nu }\overline{\Psi }_{2}-2(\overline{\gamma }+2\overline{\mu })%
\overline{\Psi }_{3}+(4\overline{\beta }-\overline{\tau })\overline{\Psi }%
_{4}.  \notag
\end{eqnarray}

The $\lambda _{i}^{a}\equiv \ $($l^{a},n^{a},m^{a},\overline{m}^{a}$) are
the components of a null tetrad, the $\Psi \ $'s and $\overline{\Psi }\ $'s
are the components of the Weyl tensor, $\lambda _{i}^{a}\nabla
_{a}=(D,\Delta ,\delta ,\overline{\delta })\ $are the directional
derivatives. \ All the other variables are the spin-coefficients. \ The
Einstein equations are already built into the system by virtue of the Ricci
tensor having been set to zero.

\section{III. \ Left-Flat Spin-Coefficient Version of Einstein equations}

There are now several different conditions that are now imposed on these
equations to restrict them to left flat equations and in addition to greatly
simplify them. \ 

The first - and most basic - is to impose the left-flat (or self-dual)
condition on the equations. This is accomplished by simply taking%
\begin{equation}
(\Psi _{0},\Psi _{1},\Psi _{2},\Psi _{3},\Psi _{4})\equiv 0,  \label{SD}
\end{equation}%
which is equivalent to imposing on the Weyl tensor $C_{abcd}\ $that

\begin{equation}
C_{abcd}\epsilon ^{cdef}=iC_{ab}^{.\ .\ ef}.  \label{SDa}
\end{equation}

All the barred variables (e.g., $\overline{\sigma },\overline{\lambda }....$%
) are independent and freed from their dual counterparts. This is denoted by
replacing the bars by tildes, (e.g., $\widetilde{\sigma },\widetilde{\lambda 
}...$).

The other conditions (both coordinate and tetrad conditions) are imposed on
the tetrad vectors, $\lambda _{i}^{a}.\ \ $We choose a world-line, $\emph{%
l,\ }$in the complex thickened $R^{4}\ $and a one parameter family of
(complex) light cones with apex on $\emph{l\ }$labeled by the complex
coordinate $x^{0}=u$. The null generators of the cones (null geodesics) are
labeled by the points of the thickened sphere with complex stereographic
coordinates $x^{A}\ $=\ ($\zeta ,\widetilde{\zeta }$), with $\widetilde{%
\zeta }\approx \overline{\zeta }.\ $The affine parameter of the null
geodesics is the complex radial coordinate $x^{1}=r.\ $These conditions
allow us$^{18}$ to write the directional derivatives (or tetrad) as%
\begin{eqnarray}
D &=&l^{a}\frac{\partial }{\partial x^{a}}=\frac{\partial }{\partial r}
\label{tet1} \\
\nabla &=&n^{a}\frac{\partial }{\partial x^{a}}=\frac{\partial }{\partial u}%
+U\frac{\partial }{\partial r}+X^{A}\frac{\partial }{\partial x^{A}},~\
(x^{2},x^{3})=(\zeta ,\widetilde{\zeta }),  \label{tet2} \\
\delta &=&m^{a}\frac{\partial }{\partial x^{a}}=\omega \frac{\partial }{%
\partial r}+\xi ^{A}\frac{\partial }{\partial x^{A}},  \label{tet3} \\
\widetilde{\delta } &=&\widetilde{m}^{a}\frac{\partial }{\partial x^{a}}=%
\widetilde{\omega }\frac{\partial }{\partial r}+\widetilde{\xi }^{A}\frac{%
\partial }{\partial x^{A}},  \label{tet4}
\end{eqnarray}%
and associated metric, [$g^{ab}=l^{a}n^{b}+n^{a}l^{b}-m^{a}\widetilde{m}^{b}-%
\widetilde{m}^{a}m^{b}$]$,$%
\begin{equation}
g^{ab}=\left[ 
\begin{array}{ccc}
0\ , & 1\ , & 0 \\ 
1\ , & g^{11}\ , & g^{1A} \\ 
0\ , & g^{1A}\ , & g^{AB}%
\end{array}%
\right]  \label{g}
\end{equation}%
with 
\begin{eqnarray}
g^{22} &=&2(U-\omega \widetilde{\omega }),  \label{gab} \\
g^{2A} &=&X^{A}-(\widetilde{\omega }\xi ^{A}+\omega \widetilde{\xi }^{A}), 
\notag \\
g^{AB} &=&-(\xi ^{A}\widetilde{\xi }^{B}+\widetilde{\xi }^{A}\xi ^{B}). 
\notag
\end{eqnarray}%
\ \ \qquad

By appropriate tetrad transformations$^{18}$,\ (parallel propagation of $%
n^{a},m^{a},\widetilde{m}^{a}$ and appropriate scaling of $l^{a}$) we can
put the following conditions on the spin-coefficients:%
\begin{equation}
\kappa =\pi =\epsilon =\widetilde{\kappa }=\widetilde{\pi }=\widetilde{%
\epsilon }=\rho -\widetilde{\rho }=0.  \label{tetconditions}
\end{equation}

We also have, but do not use immediately,\qquad 
\begin{equation}
\ \tau =\widetilde{\alpha }+\beta ,\ \ \ \widetilde{\tau }=\alpha +%
\widetilde{\beta }.
\end{equation}

These conditions are then inserted into our Einstein field equations to
obtain our final set of field equations. The equations are displayed in
groups: first, (17) of them, that contain the $D\ $derivatives, then, (7),
containing only the angular derivatives, ($\delta ,\widetilde{\delta }$ )
and finally, (20), containing the time derivative, $\Delta .$

\begin{eqnarray}
D\rho &=&\rho ^{2}+\sigma \widetilde{\sigma }  \label{D1} \\
D\sigma &=&2\rho \sigma ,\ \ \ \ \ \ \ \ \ \ \   \label{D2} \\
D\widetilde{\sigma } &=&2\rho \widetilde{\sigma }+\widetilde{\Psi }_{0}
\label{D3} \\
D\tau &=&\tau \rho +\widetilde{\tau }\sigma ,\ \ \   \label{D4} \\
\ D\widetilde{\tau } &=&\widetilde{\tau }\rho +\tau \widetilde{\sigma }+%
\widetilde{\Psi }_{1}  \label{D5} \\
D\alpha &=&\rho \alpha +\beta \widetilde{\sigma }  \label{D6} \\
D\widetilde{\alpha } &=&\rho \widetilde{\alpha }+\widetilde{\beta }\sigma
\label{D7} \\
D\beta &=&\alpha \sigma +\rho \beta  \label{D8} \\
D\widetilde{\beta } &=&\widetilde{\alpha }\widetilde{\sigma }+\rho 
\widetilde{\beta }+\widetilde{\Psi }_{1}  \label{D9} \\
D\gamma &=&\tau \alpha +\widetilde{\tau }\beta  \label{D10} \\
D\widetilde{\gamma } &=&\widetilde{\tau }\widetilde{\alpha }+\tau \widetilde{%
\beta }+\widetilde{\Psi }_{2}  \label{D11} \\
D\lambda &=&\rho \lambda +\widetilde{\sigma }\mu  \label{D12} \\
D\widetilde{\lambda } &=&\rho \widetilde{\lambda }+\sigma \widetilde{\mu }
\label{D13} \\
D\mu &=&\rho \mu +\sigma \lambda  \label{D14} \\
D\widetilde{\mu } &=&\rho \widetilde{\mu }+\widetilde{\sigma }\widetilde{%
\lambda }+\widetilde{\Psi }_{2}  \label{D15} \\
D\nu &=&\widetilde{\tau }\mu +\tau \lambda  \label{D16} \\
D\widetilde{\nu } &=&\tau \widetilde{\mu }+\widetilde{\tau }\widetilde{%
\lambda }+\widetilde{\Psi }_{3}  \label{D17a}
\end{eqnarray}

\begin{eqnarray}
DU &=&\tau \widetilde{\omega }+\widetilde{\tau }\omega -(\gamma +\widetilde{%
\gamma }),  \label{m1} \\
DX^{A} &=&\tau \widetilde{\xi }^{A}+\widetilde{\tau }\xi ^{A},  \label{m2} \\
D\omega  &=&\sigma \widetilde{\omega }+\rho \omega -\tau ,  \label{m3} \\
D\widetilde{\omega } &=&\widetilde{\sigma }\omega +\rho \widetilde{\omega }-%
\widetilde{\tau },  \label{m4a} \\
D\xi ^{A} &=&\sigma \widetilde{\xi }^{A}+\rho \xi ^{A},  \label{m5} \\
D\widetilde{\xi }^{A} &=&\widetilde{\sigma }\xi ^{A}+\rho \widetilde{\xi }%
^{A},  \label{m6}
\end{eqnarray}

\begin{subequations}
\begin{eqnarray}
\delta \widetilde{\Psi }_{0}-D\widetilde{\Psi }_{1} &=&4\widetilde{\alpha }%
\widetilde{\Psi }_{0}-4\rho \widetilde{\Psi }_{1},  \label{BI1} \\
\delta \widetilde{\Psi }_{1}-D\widetilde{\Psi }_{2} &=&\widetilde{\lambda }%
\widetilde{\Psi }_{0}+2\widetilde{\alpha }\widetilde{\Psi }_{1}-3\rho 
\widetilde{\Psi }_{2},  \label{BI2} \\
\delta \widetilde{\Psi }_{2}-D\widetilde{\Psi }_{3} &=&2\widetilde{\lambda }%
\widetilde{\Psi }_{1}-2\rho \widetilde{\Psi }_{3},  \label{BI3} \\
\delta \widetilde{\Psi }_{3}-D\widetilde{\Psi }_{4} &=&3\widetilde{\lambda }%
\widetilde{\Psi }_{2}-2\widetilde{\alpha }\widetilde{\Psi }_{3}-\rho 
\widetilde{\Psi }_{4},  \label{BI4}
\end{eqnarray}

\end{subequations}
\begin{eqnarray}
\delta \rho -\widetilde{\delta }\sigma &=&\rho (\widetilde{\alpha }+\beta
)-\sigma (3\alpha -\widetilde{\beta })  \label{m7} \\
\widetilde{\delta }\rho -\delta \widetilde{\sigma } &=&\rho (\alpha +%
\widetilde{\beta })-\widetilde{\sigma }(3\widetilde{\alpha }-\beta )-%
\widetilde{\Psi }_{1}  \label{m8} \\
\delta \alpha -\widetilde{\delta }\beta &=&\mu \rho -\lambda \sigma +\alpha 
\widetilde{\alpha }+\beta \widetilde{\beta }-2\alpha \beta  \label{m9} \\
\widetilde{\delta }\widetilde{\alpha }-\delta \widetilde{\beta } &=&%
\widetilde{\mu }\rho -\widetilde{\lambda }\widetilde{\sigma }+\alpha 
\widetilde{\alpha }+\beta \widetilde{\beta }-2\widetilde{\alpha }\widetilde{%
\beta }-\widetilde{\Psi }_{2}  \label{m10} \\
\delta \lambda -\widetilde{\delta }\mu &=&\mu (\alpha +\widetilde{\beta }%
)+\lambda (\widetilde{\alpha }-3\beta )  \label{m11} \\
\widetilde{\delta }\widetilde{\lambda }-\delta \widetilde{\mu } &=&%
\widetilde{\mu }(\widetilde{\alpha }+\beta )+\widetilde{\lambda }(\alpha -3%
\widetilde{\beta })-\widetilde{\Psi }_{3}  \label{m12}
\end{eqnarray}

\begin{eqnarray}
\widetilde{\delta }\omega -\delta \widetilde{\omega } &=&(\widetilde{\mu }%
-\mu )-(\widetilde{\alpha }-\beta )\widetilde{\omega }+(\alpha -\widetilde{%
\beta })\omega .  \label{m14} \\
\widetilde{\delta }\xi ^{A}-\delta \widetilde{\xi }^{A} &=&-(\widetilde{%
\alpha }-\beta )\widetilde{\xi }^{A}+(\alpha -\widetilde{\beta })\xi ^{A}
\label{m15}
\end{eqnarray}%
\begin{eqnarray}
\delta U-\Delta \omega &=&-\nu +\widetilde{\lambda }\widetilde{\omega }+(\mu
-\gamma +\widetilde{\gamma })\omega ,  \label{t1} \\
\widetilde{\delta }U-\Delta \widetilde{\omega } &=&-\widetilde{\nu }+\lambda
\omega +(\widetilde{\mu }-\widetilde{\gamma }+\gamma )\widetilde{\omega },
\label{t2} \\
\delta X^{A}-\Delta \xi ^{A} &=&\widetilde{\lambda }\widetilde{\xi }%
^{A}+(\mu -\gamma +\widetilde{\gamma })\xi ^{A}  \label{t3} \\
\widetilde{\delta }X^{A}-\Delta \widetilde{\xi }^{A} &=&\lambda \xi ^{A}+(%
\widetilde{\mu }-\widetilde{\gamma }+\gamma )\widetilde{\xi }^{A}  \label{t4}
\end{eqnarray}

\begin{eqnarray}
\Delta \lambda -\widetilde{\delta }\nu &=&-(\mu +\widetilde{\mu })\lambda
-(3\gamma -\widetilde{\gamma })+2\alpha \lambda  \label{ts1} \\
\Delta \widetilde{\lambda }-\delta \widetilde{\nu } &=&-(\mu +\widetilde{\mu 
})\widetilde{\lambda }-(3\widetilde{\gamma }-\gamma )+2\widetilde{\alpha }%
\widetilde{\lambda }-\widetilde{\Psi }_{4}  \label{ts2} \\
\delta \nu -\Delta \mu &=&\mu ^{2}+\lambda \widetilde{\lambda }+\mu (\gamma +%
\widetilde{\gamma })-2\beta \nu  \label{ts3} \\
\widetilde{\delta }\widetilde{\nu }-\Delta \widetilde{\mu } &=&\widetilde{%
\mu }^{2}+\lambda \widetilde{\lambda }+\widetilde{\mu }(\gamma +\widetilde{%
\gamma })-2\widetilde{\beta }\widetilde{\nu }  \label{ts4} \\
\delta \gamma -\Delta \beta &=&\mu \tau -\sigma \nu -\beta (\gamma -%
\widetilde{\gamma }-\mu )+\alpha \widetilde{\lambda }  \label{ts5} \\
\widetilde{\delta }\widetilde{\gamma }-\Delta \widetilde{\beta } &=&%
\widetilde{\mu }\widetilde{\tau }-\widetilde{\sigma }\widetilde{\nu }-%
\widetilde{\beta }(\widetilde{\gamma }-\gamma -\widetilde{\mu })+\widetilde{%
\alpha }\lambda  \label{ts6} \\
\delta \tau -\Delta \sigma &=&(\mu \sigma +\rho \widetilde{\lambda })+\tau
(\tau +\beta -\widetilde{\alpha })-\sigma (3\gamma -\widetilde{\gamma })
\label{ts7} \\
\widetilde{\delta }\widetilde{\tau }-\Delta \widetilde{\sigma } &=&(%
\widetilde{\mu }\widetilde{\sigma }+\rho \lambda )+\widetilde{\tau }(%
\widetilde{\tau }+\widetilde{\beta }-\alpha )-\widetilde{\sigma }(3%
\widetilde{\gamma }-\gamma )  \label{ts8} \\
\Delta \rho -\widetilde{\delta }\tau &=&-(\rho \widetilde{\mu }+\sigma
\lambda )+\tau (\widetilde{\beta }-\alpha -\widetilde{\tau })+(\gamma +%
\widetilde{\gamma })\rho  \label{ts9} \\
\Delta \rho -\delta \widetilde{\tau } &=&-(\rho \mu +\widetilde{\sigma }%
\widetilde{\lambda })+\widetilde{\tau }(\beta -\widetilde{\alpha }-\tau
)+(\gamma +\widetilde{\gamma })\rho -\widetilde{\Psi }_{2}  \label{ts10} \\
\Delta \alpha -\widetilde{\delta }\gamma &=&\nu \rho -\lambda (\tau +\beta
)+\alpha (\widetilde{\gamma }-\widetilde{\mu })+\gamma (\widetilde{\beta }-%
\widetilde{\tau })  \label{ts11} \\
\Delta \widetilde{\alpha }-\delta \widetilde{\gamma } &=&\widetilde{\nu }%
\rho -\widetilde{\lambda }(\widetilde{\tau }+\widetilde{\beta })+\widetilde{%
\alpha }(\gamma -\mu )+\widetilde{\gamma }(\beta -\tau )-\widetilde{\Psi }%
_{3}  \label{ts12}
\end{eqnarray}

\begin{eqnarray}
\Delta \widetilde{\Psi }_{0}-\widetilde{\delta }\widetilde{\Psi }_{1} &=&(4%
\widetilde{\gamma }-\widetilde{\mu })\widetilde{\Psi }_{0}-2(2\widetilde{%
\tau }+\widetilde{\beta })\widetilde{\Psi }_{1}+3\widetilde{\sigma }%
\widetilde{\Psi _{2}},  \label{BI5} \\
\Delta \widetilde{\Psi }_{1}-\widetilde{\delta }\widetilde{\Psi }_{2} &=&%
\widetilde{\nu }\widetilde{\Psi }_{0}+2(\widetilde{\gamma }-\widetilde{\mu })%
\widetilde{\Psi }_{1}-3\widetilde{\tau }\widetilde{\Psi }_{2}+2\widetilde{%
\sigma }\widetilde{\Psi }_{3},  \label{BI6} \\
\Delta \widetilde{\Psi }_{2}-\widetilde{\delta }\widetilde{\Psi }_{3} &=&2%
\widetilde{\nu }\widetilde{\Psi }_{1}-3\widetilde{\mu }\widetilde{\Psi }%
_{2}+2(\widetilde{\beta }-\widetilde{\tau })\widetilde{\Psi }_{3}+\widetilde{%
\sigma }\widetilde{\Psi }_{4},  \label{BI7} \\
\Delta \widetilde{\Psi }_{3}-\widetilde{\delta }\widetilde{\Psi }_{4} &=&3%
\widetilde{\nu }\widetilde{\Psi }_{2}-2(\widetilde{\gamma }+2\widetilde{\mu }%
)\widetilde{\Psi }_{3}+(4\widetilde{\beta }-\widetilde{\tau })\widetilde{%
\Psi }_{4}.  \label{BI8}
\end{eqnarray}

In the following section we integrate explicitly all the $D$ equations. The
solution to each equation will have an associated 'constant' (actually a
function independent of $r$) of integration. These `constants' are
determined by either the angular equations or by the fact of all the null
surfaces being light-cones with their origin on a world-line. \ The
evolution equations (i.e., those with $\Delta $) turn out - initially - to
be extremely complicated to deal with. Nevertheless - and quite surprising -
in the end they all are equivalent to one single non-linear wave equation.

\section{IV. Integration}

\subsection{Radial Integration}

\subsubsection{preliminary integration}

The first thing to note is that Eqs.(\ref{D1}) and (\ref{D2}) can be
integrated immediately, (with the coordinate origin for $r\ \ $taken at the
apex of the light-cone) as%
\begin{eqnarray}
\rho  &=&-\frac{1}{r},  \label{rho} \\
\sigma  &=&0.  \label{sigma}
\end{eqnarray}%
The later follows from the fact that $\sigma \ $vanishes at a light-cone
apex.

The field equations are greatly simplified by these results. The
spin-coefficient Eqs.(\ref{D4},\ref{D7}, \ref{D8},

\noindent \ref{D13}),\ref{D14} decouple from the remainder and using the
results from flat space light-cones with origins on a world-line$^{1}$\ we
have%
\begin{eqnarray}
\tau  &=&\overline{\alpha }+\beta =0,  \label{tau} \\
\overline{\alpha } &=&r^{-1}\widetilde{\alpha }^{0}=-\frac{1}{2}%
r^{-1}\partial P,\ \ \ P=1+\zeta \widetilde{\zeta },  \label{alphbar} \\
\beta  &=&-\widetilde{\alpha },  \label{beta} \\
\widetilde{\lambda } &=&0,  \label{lambdabar} \\
\mu  &=&-r^{-1}.  \label{mu}
\end{eqnarray}

The form of the function $P,\ $(\ref{alphbar})$,\ $which determines the
2-sphere metric at the cones apex via Eqs.(\ref{gab}), is set as a
coordinate condition.

In a similar fashion, with $\tau =0,\ $the metric Eqs.(\ref{m3}), (\ref{m5})
integrate to%
\begin{eqnarray}
\xi ^{A} &=&r^{-1}(\xi ^{0\zeta },\xi ^{0\widetilde{\zeta }})=-r^{-1}(P,0),
\label{xi} \\
\omega  &=&0.  \label{omega}
\end{eqnarray}%
$\ $

\subsubsection{ The Radial Bianchi Identities \& Weyl Tensor}

Turning now to our first non-trivial integrations, using the preceding
results, the four radial Bianchi Identities, (\ref{BI1})-\ref{BI4}) become

\begin{subequations}
\begin{eqnarray}
\partial _{r}\widetilde{\Psi }_{1} &=&-4r^{-1}\widetilde{\Psi }%
_{1}-r^{-1}\eth \widetilde{\Psi }_{0},  \label{BIA1} \\
\partial _{r}\widetilde{\Psi }_{2} &=&-3r^{-1}\widetilde{\Psi }%
_{2}-r^{-1}\eth \widetilde{\Psi }_{1},  \label{BIA2} \\
\partial _{r}\widetilde{\Psi }_{3} &=&-2r^{-1}\widetilde{\Psi }%
_{3}-r^{-1}\eth \widetilde{\Psi }_{2},  \label{BIA3} \\
\partial _{r}\widetilde{\Psi }_{4} &=&-r^{-1}\widetilde{\Psi }%
_{4}-r^{-1}\eth \widetilde{\Psi }_{3},  \label{BIA4}
\end{eqnarray}

\noindent where Eqs.(\ref{tet1}) and (\ref{tet4}), along with $\eth \eta
_{s}=P^{1-s}\frac{\partial }{\partial \zeta }(P^{s}\eta _{s}),\ $were used.$%
\ $

Writing these equations, with $n=1,2,3,4$, succinctly as 
\end{subequations}
\begin{equation}
\partial _{r}\widetilde{\Psi }_{5-n}=-nr^{-1}\widetilde{\Psi }%
_{5-n}-r^{-1}\eth \widetilde{\Psi }_{4-n}  \label{BIA*}
\end{equation}
the solutions are given by

\begin{equation}
\widetilde{\Psi }_{5-n}=r^{-n}\widetilde{\Psi }_{5-n}^{(0)}-r^{-n}\eth
\int^{r}r^{n-1}\widetilde{\Psi }_{4-n}dr,  \label{sol.1}
\end{equation}%
the integral being indefinite, with $\widetilde{\Psi }_{5-n}^{(0)}\ $the
"constants of integration".

Our integration process continues with the introduction of the set of five,
spin-weight $s=-2,\ $potentials, $\check{\Upsilon}_{5-n}(r)$, ($n=1,2,3,4,5$)%
$\ $by

\begin{equation}
\widetilde{\Psi }_{5-n}=(-1)^{n-1}r^{-n+2}\eth ^{(5-n)}\partial _{r}\check{%
\Upsilon}_{6-n}  \label{potentials.1}
\end{equation}

\noindent and inserting them into the integral of Eq.(\ref{sol.1}).

Integration leads to

\begin{equation}
\widetilde{\Psi }_{5-n}=r^{-n}\widetilde{\Psi }%
_{5-n(0)}^{(0)}-(-1)^{n}r^{-n}\eth ^{(5-n)}\check{\Upsilon}_{5-n}(r),
\label{sol.2}
\end{equation}%
where, from regularity at $r=0,\ $we have%
\begin{equation}
\widetilde{\Psi }_{5-n(0)}^{(0)}=0.  \label{no singularity}
\end{equation}

\begin{remark}
We emphasize that there are two different types of situations. In one case
the potentials, $\check{\Upsilon},$ vanish sufficiently fast at $r=0\ $so
that the origin is a regular point of the manifold. In the other case there
are intrinsic singularities hidden in the $\check{\Upsilon}(r)\ $at\ $r=0.\ $%
See the following section for examples.]\qquad \qquad
\end{remark}

Written explicitly, Eqs.(\ref{sol.2}) are%
\begin{eqnarray}
&&\widetilde{\Psi }_{0}=r^{-5}\check{\Upsilon}_{0},  \label{sol.a} \\
&&\widetilde{\Psi }_{1}=-r^{-4}\eth \check{\Upsilon}_{1},  \label{sol.b} \\
\text{\ \ \ } &&\widetilde{\Psi }_{2}=r^{-3}\eth ^{2}\check{\Upsilon}_{2},
\label{sol.c} \\
\text{\ \ \ } &&\widetilde{\Psi }_{3}=-r^{-2}\eth ^{3}\check{\Upsilon}_{3},
\label{sol.d} \\
&&\widetilde{\Psi }_{4}=r^{-1}\eth ^{4}\check{\Upsilon}_{4}.  \label{sol.e}
\end{eqnarray}

By equating Eqs.(\ref{potentials.1}) and (\ref{sol.2}), we find the `ladder'
relationship between the different potentials,%
\begin{equation}
\check{\Upsilon}_{5-n}=r^{2}\partial _{r}\check{\Upsilon}_{6-n}
\label{g.ladder}
\end{equation}

\noindent or%
\begin{eqnarray}
\text{\ } &&\check{\Upsilon}_{0}=r^{2}\partial _{r}\check{\Upsilon}_{1},
\label{g0} \\
&&\check{\Upsilon}_{1}=r^{2}\partial _{r}\check{\Upsilon}_{2},  \label{g1} \\
&&\check{\Upsilon}_{2}=r^{2}\partial _{r}\check{\Upsilon}_{3},  \label{g2} \\
&&\check{\Upsilon}_{3}=r^{2}\partial _{r}\check{\Upsilon}_{4}.  \label{g3}
\end{eqnarray}

Knowledge of $\check{\Upsilon}_{4}\ $(which we can consider as free initial
data)$\ $determines the others or equivalently, from $\check{\Upsilon}_{0}\ $%
with the four constants of integration, from going up the ladder to $\check{%
\Upsilon}_{4}$, determines the others.

In the case of the regularity of the Weyl components at $r=0,\ $we must
impose conditions on the\ $\check{\Upsilon}_{n}\ $so that near $r=0\ $

\begin{equation}
\check{\Upsilon}_{n}=O(r^{5-n}).  \label{O(5)}
\end{equation}

\noindent This, with suitable differentiability, follows from $\check{%
\Upsilon}_{4}=Ar+O(r^{2}).$

For conventional asymptotic flatness (peeling) we require that, as $%
r\rightarrow \infty ,\ $all the potentials tend to a constant, $\check{%
\Upsilon}_{n}=c_{n}+O(r^{-1}).$

From Eq.(\ref{g.ladder}) we have the asymptotic behavior of the potentials:

\begin{eqnarray}
\check{\Upsilon}_{0} &=&r^{2}\partial _{r}\check{\Upsilon}%
_{1}=c_{0}+0(r^{-1}),  \label{asy.0} \\
\check{\Upsilon}_{1} &=&r^{2}\partial _{r}\check{\Upsilon}%
_{2}=c_{1}-c_{0}r^{-1}+0(r^{-2}),  \label{asy.1} \\
\check{\Upsilon}_{2} &=&r^{2}\partial _{r}\check{\Upsilon}%
_{3}=c_{2}-c_{1}r^{-1}+\frac{1}{2}c_{0}r^{-2}+0(r^{-3}),  \label{asy.2} \\
\check{\Upsilon}_{3} &=&r^{2}\partial _{r}\check{\Upsilon}%
_{4}=c_{3}-c_{2}r^{-1}+\frac{1}{2}c_{1}r^{-2}-\frac{1}{6}%
c_{0}r^{-3}+0(r^{-4}),  \label{asym3} \\
\check{\Upsilon}_{4} &=&c_{4}-c_{3}r^{-1}+\frac{1}{2}c_{2}r^{-2}-\frac{1}{6}%
c_{1}r^{-3}+\frac{1}{24}c_{0}r^{-4}+0(r^{-5}).  \label{asy.4}
\end{eqnarray}

\subsubsection{Remaining Radial Equations}

The remaining spin-coefficient and metric equations can be integrated and
expressed in terms of the potentials.

We first investigate Eq.(\ref{D3}) using $\widetilde{\Psi }_{0}=r^{-5}\check{%
\Upsilon}_{0},\ $i.e., 
\begin{equation}
D\widetilde{\sigma }=-2r^{-1}\widetilde{\sigma }+r^{-5}\check{\Upsilon}_{0}.
\label{D.3*}
\end{equation}

\noindent Its solution is given by

\begin{equation}
\widetilde{\sigma }=\widetilde{\sigma }^{0}r^{-2}+r^{-2}\int^{r}r^{-3}\check{%
\Upsilon}_{0}dr.  \label{sigma twiddle}
\end{equation}%
with (again) an indefinite integral and $\widetilde{\sigma }^{0}\ $being the
constant of integration.

The integral term can be greatly simplified via repeated use of Eq.(\ref%
{g.ladder}). \ From

\begin{eqnarray}
\int^{r}r^{-3}\check{\Upsilon}_{0}dr &=&\int^{r}r^{-1}\partial _{r}\check{%
\Upsilon}_{1}dr=\int^{r}[\partial _{r}(r^{-1}\check{\Upsilon}_{1})+r^{-2}%
\check{\Upsilon}_{1}]dr  \label{manipulate.1} \\
&=&r^{-1}\check{\Upsilon}_{1}+\int^{r}r^{-2}\check{\Upsilon}_{1}dr=r^{-1}%
\check{\Upsilon}_{1}+\int^{r}\partial _{r}\check{\Upsilon}_{2}dr  \notag \\
&=&r^{-1}\check{\Upsilon}_{1}+\check{\Upsilon}_{2}  \notag
\end{eqnarray}

\noindent we have%
\begin{equation}
\widetilde{\sigma }=\widetilde{\sigma }^{0}r^{-2}+r^{-3}\check{\Upsilon}%
_{1}+r^{-2}\check{\Upsilon}_{2}
\end{equation}

From the vanishing of the $\check{\Upsilon}\ $at $r=0\ $and the regularity
of the light-cones at the origin, we take $\widetilde{\sigma }^{0}=0,\ $\ so
that

\begin{equation}
\widetilde{\sigma }=r^{-3}\check{\Upsilon}_{1}+r^{-2}\check{\Upsilon}_{2}.
\label{sigma twiddle.2}
\end{equation}

Using the same methods, i.e., the formal radial integration of the equations
followed by the repeated use of the ladder relations, Eq.(\ref{g.ladder}),
all the remaining radial equations can be integrated and expressed in terms
of the $\check{\Upsilon}_{n}$. \ Many of the associated "constants of
integration" are determined by the behavior near $r=0$,\ $\ $several are
determined by use of the angular equation. (It is likely that even those
determined from the angular equations could have been determined from their $%
r=0\ $behavior.)

The following is the full set of solutions to all the spin-coefficient and
metric radial equations.

Spin-coefficients:%
\begin{eqnarray}
\ \ \sigma &=&\tau =\widetilde{\alpha }+\beta =0,\ \ \ \widetilde{\tau }%
=\alpha +\ \widetilde{\beta },  \label{sc1} \\
\ \ \rho &=&\widetilde{\rho }=-r^{-1},\ \ \ \ \ \ \ \ \   \label{sc3} \\
\ \widetilde{\sigma } &=&r^{-2}\check{\Upsilon}_{2}+r^{-3}\ \check{\Upsilon}%
_{1},  \label{sc4} \\
\ \widetilde{\tau } &=&-r^{-2}\ \eth \check{\Upsilon}_{2}-r^{-1}\eth \check{%
\Upsilon}_{3}\ \ \ ,  \label{sc5} \\
\ \widetilde{\alpha } &=&r^{-1}\widetilde{\alpha }^{0}=-\frac{1}{2}%
r^{-1}\partial _{\zeta }P,  \label{sc6} \\
\ \beta &=&r^{-1}\beta ^{0}=-\widetilde{\alpha },\ \ \ \ \ \   \label{sc7} \\
\alpha &=&-r^{-1}(\frac{1}{2}\partial _{\widetilde{\zeta }}P-\partial
_{\zeta }P\check{\Upsilon}_{3})+r^{-2}\ \frac{1}{2}\partial _{\zeta }P\check{%
\Upsilon}_{2},  \label{sc8} \\
\ \widetilde{\beta } &=&r^{-1}\{\frac{1}{2}\partial _{\widetilde{\zeta }%
}P-\eth \check{\Upsilon}_{3}-\partial _{\zeta }P\check{\Upsilon}%
_{3}\}-r^{-2}\ (\eth \check{\Upsilon}_{2}+\frac{1}{2}\partial _{\zeta }P%
\check{\Upsilon}_{2})  \label{sc9} \\
\gamma &=&-\partial _{\zeta }P\eth \check{\Upsilon}_{4}-r^{-1}\frac{1}{2}%
\partial _{\zeta }P\eth \check{\Upsilon}_{3}  \label{sc10} \\
\ \widetilde{\gamma } &=&\eth ^{2}\check{\Upsilon}_{4}+\partial _{\zeta
}P\eth \check{\Upsilon}_{4}+r^{-1}(\eth ^{2}\check{\Upsilon}_{3}+\frac{1}{2}%
\partial _{\zeta }P\eth \check{\Upsilon}_{3})  \label{sc11} \\
\ \widetilde{\lambda } &=&0  \label{sc12} \\
\ \mu &=&-r^{-1}  \label{sc13} \\
\ \lambda &=&-\ 2r^{-1}\check{\Upsilon}_{3}-r^{-2}\check{\Upsilon}_{2}
\label{sc14} \\
\ \widetilde{\mu } &=&-r^{-1}+r^{-1}\eth ^{2}\check{\Upsilon}_{3}\ 
\label{sc15} \\
\ \ \widetilde{\nu } &=&-\eth ^{3}\check{\Upsilon}_{4}  \label{sc16} \\
\ \nu &=&2\eth \check{\Upsilon}_{4}+r^{-1}\eth \check{\Upsilon}_{3}
\label{sc17}
\end{eqnarray}

Metric Variables:%
\begin{eqnarray}
\ \xi ^{A} &=&(\xi ^{\zeta },\xi ^{\widetilde{\zeta }})=r^{-1}\xi
^{0A}=-r^{-1}(P,0),\ \ \ \ P=1+\zeta \widetilde{\zeta }  \label{MV} \\
\ \widetilde{\xi }^{A} &=&(\widetilde{\xi }^{\zeta },\widetilde{\xi }^{%
\widetilde{\zeta }})=-r^{-1}(0,P)-(P,0)(2r^{-1}\check{\Upsilon}_{3}+r^{-2}%
\check{\Upsilon}_{2})  \label{mv3} \\
\ \omega &=&r^{-1}\omega ^{0}=0  \label{mv4} \\
\ \widetilde{\omega } &=&r^{-1}\widetilde{\omega }^{0}+\eth \check{\Upsilon}%
_{3}=\eth \check{\Upsilon}_{3}  \label{mv5} \\
\ X^{A} &=&(X^{\zeta },X^{\widetilde{\zeta }})=(P,0)[r^{-1}\eth \check{%
\Upsilon}_{3}+2\eth \check{\Upsilon}_{4}]  \label{mv6} \\
U &=&-1-r\eth ^{2}\check{\Upsilon}_{4}  \label{mv7}
\end{eqnarray}

From these results we have the derivative operators which are needed in the
angular and evolution equations, namely%
\begin{eqnarray}
D &=&\partial _{r},  \label{0.1} \\
\nabla &=&\partial _{u}-(1+r\eth ^{2}\check{\Upsilon}_{4})\partial
_{r}+(r^{-1}\ P\eth \check{\Upsilon}_{3}+2P\eth \check{\Upsilon}%
_{4})\partial _{\zeta },  \label{0.2} \\
\delta &=&-r^{-1}P\partial _{\zeta },  \label{0.3} \\
\overline{\delta } &=&\eth \check{\Upsilon}_{3}\partial _{r}-r^{-1}P\partial 
\widetilde{_{\zeta }}-P(r^{-2}\check{\Upsilon}_{2}+2r^{-1}\Upsilon
_{2})\partial _{\zeta }.  \label{0.4}
\end{eqnarray}

\subsection{Angular Equations}

Virtually all the angular derivative spin-coefficient equations turn out to
be identities when the results of the previous section are used. \ Some
determine the "constants of integration" but they are almost certainly also
determined by the behavior near $r=0.\ $Several examples, (already used in
the previous subsection) are:

\begin{eqnarray}
\ \delta X^{A}-\Delta \xi ^{A} &=&\xi ^{A}(\mu +\widetilde{\gamma }-\gamma )+%
\widetilde{\lambda }\widetilde{\xi }^{A}=>\gamma ^{0}=0, \\
\delta \widetilde{\xi }^{A}-\widetilde{\delta }\xi ^{A} &=&(\widetilde{\beta 
}-\alpha )\xi ^{A}+(\widetilde{\alpha }-\beta )\widetilde{\xi }^{A}=>%
\widetilde{\alpha }^{0}=-\frac{1}{2}\partial _{\zeta }P, \\
\delta \widetilde{\omega }-\widetilde{\delta }\omega &=&(\widetilde{\beta }%
-\alpha )\omega +(\widetilde{\alpha }-\beta )\widetilde{\omega }+(\mu -%
\widetilde{\mu })=>Identity, \\
\delta U-\Delta \omega &=&(\mu +\widetilde{\gamma }-\gamma )\omega +%
\widetilde{\lambda }\widetilde{\omega }-\widetilde{\nu }=>\ \widetilde{\nu }%
^{0}=0.
\end{eqnarray}

\subsection{Evolution Equations}

Analyzing the twenty evolution equations for the five potentials, Eqs.(\ref%
{t1})-(\ref{BI8}), initially presented a very difficult challenge. \ The
process of inserting all the previous results, i.e., Eqs.(\ref{sol.a})-(\ref%
{sol.e}) and Eqs.(\ref{sc1})-(\ref{mv7}), into the evolution equations was a
daunting task. Analyzing any one could take several days. The algebra was
long and complicated with the easy production of many errors. The saving
aspect of it was the fact that the equations were interrelated, in the sense
that final forms of many of the equations could be compared or reduced to
others, thereby giving a method of checking consistency and thus finding the
errors.

Towards the end of the analysis all the evolution equations depended on just
the four evolutionary Bianchi Identities,\ Eqs.(\ref{t1}) - (\ref{t4}). They
however, using Eqs.(\ref{g0}) - (\ref{g3}), are further reduced\ to a single
equation for the last of the potentials, i.e., $\check{\Upsilon}_{4}.$

Taking the four evolutionary Bianchi Identities, Eqs.(\ref{BI5})-(\ref{BI8}%
), substituting the results of the radial integrations - with lengthy
calculations using Eqs.(\ref{g0}) -(\ref{g3}), - we obtain the following
results:%
\begin{eqnarray}
&&  \label{0'} \\
&&\check{\Upsilon}_{0}^{\prime }-\widetilde{\eth }\eth \check{\Upsilon}_{1}-2%
\check{\Upsilon}_{3}\eth ^{2}\check{\Upsilon}_{1}+2\eth \check{\Upsilon}%
_{3}\eth \check{\Upsilon}_{1}-3\check{\Upsilon}_{2}\eth ^{2}\check{\Upsilon}%
_{2}+2\eth \check{\Upsilon}_{4}\eth \check{\Upsilon}_{0}+\eth ^{2}\check{%
\Upsilon}_{4}\check{\Upsilon}_{0}-\frac{\partial }{\partial r}\check{\Upsilon%
}_{0}  \notag \\
&&-r^{-1}(3\check{\Upsilon}_{0}\eth ^{2}\check{\Upsilon}_{3}-2\eth \check{%
\Upsilon}_{3}\eth \check{\Upsilon}_{0}-4\check{\Upsilon}_{0}+\check{\Upsilon}%
_{2}\eth ^{2}\check{\Upsilon}_{1}-6\eth \check{\Upsilon}_{2}\eth \check{%
\Upsilon}_{1}+3\eth ^{2}\check{\Upsilon}_{2}\check{\Upsilon}_{1})  \notag \\
&{\tiny :}&-r\eth ^{2}\check{\Upsilon}_{4}\frac{\partial }{\partial r}\check{%
\Upsilon}_{0}=0,  \notag
\end{eqnarray}

\begin{eqnarray}
&&-\eth \check{\Upsilon}_{1}^{\prime }+\widetilde{\eth }\eth ^{2}\check{%
\Upsilon}_{2}-2\eth ^{2}\check{\Upsilon}_{4}\eth \check{\Upsilon}_{1}-2\eth 
\check{\Upsilon}_{4}\eth ^{2}\check{\Upsilon}_{1}+2\check{\Upsilon}_{3}\eth
^{3}\check{\Upsilon}_{2}+2\check{\Upsilon}_{2}\eth ^{3}\check{\Upsilon}_{3}
\label{1'} \\
&&+r^{-1}(-2\eth \check{\Upsilon}_{1}-3\eth \check{\Upsilon}_{2}\eth ^{2}%
\check{\Upsilon}_{2}+\check{\Upsilon}_{2}\eth ^{3}\check{\Upsilon}_{2}+2%
\check{\Upsilon}_{1}\eth ^{3}\check{\Upsilon}_{3}-2\eth ^{2}\check{\Upsilon}%
_{1}\eth \check{\Upsilon}_{3}+\eth ^{2}\check{\Upsilon}_{4}\eth \check{%
\Upsilon}_{0}+\eth ^{3}\check{\Upsilon}_{4}\check{\Upsilon}_{0})  \notag \\
&{\tiny :}&+r^{-2}\eth \check{\Upsilon}_{0}=0,  \notag
\end{eqnarray}

\begin{eqnarray}
&&  \label{2'} \\
&&\text{\textbf{\ }}\eth ^{2}\check{\Upsilon}_{2}^{\prime }-\widetilde{\eth }%
\eth ^{3}\check{\Upsilon}_{3}-2\eth \check{\Upsilon}_{3}\eth ^{3}\check{%
\Upsilon}_{3}-2\check{\Upsilon}_{3}\eth ^{4}\check{\Upsilon}_{3}-\check{%
\Upsilon}_{2}\eth ^{4}\check{\Upsilon}_{4}+2\eth \check{\Upsilon}_{4}\eth
^{3}\check{\Upsilon}_{2}+3\eth ^{2}\check{\Upsilon}_{4}\eth ^{2}\check{%
\Upsilon}_{2}  \notag \\
&&+r^{-1}(+2\eth ^{3}\check{\Upsilon}_{2}\eth \check{\Upsilon}_{3}+3\eth ^{2}%
\check{\Upsilon}_{3}\eth ^{2}\check{\Upsilon}_{2}-\check{\Upsilon}_{2}\eth
^{4}\check{\Upsilon}_{3}-\eth ^{4}\check{\Upsilon}_{4}\check{\Upsilon}%
_{1}-2\eth ^{3}\check{\Upsilon}_{4}\eth \check{\Upsilon}_{1}-\eth ^{2}\check{%
\Upsilon}_{4}\eth ^{2}\check{\Upsilon}_{1})  \notag \\
&{\tiny :}&-r^{-2}\eth ^{2}\check{\Upsilon}_{1}=0,  \notag
\end{eqnarray}

\begin{eqnarray}
&&-\eth ^{3}\check{\Upsilon}_{3}^{\prime }+\widetilde{\eth }\eth ^{4}\check{%
\Upsilon}_{4}-2\eth \check{\Upsilon}_{4}\eth ^{4}\check{\Upsilon}_{3}-4\eth
^{2}\check{\Upsilon}_{4}\eth ^{3}\check{\Upsilon}_{3}+4\eth ^{4}\check{%
\Upsilon}_{4}\eth \check{\Upsilon}_{3}+2\eth ^{5}\check{\Upsilon}_{4}\check{%
\Upsilon}_{3}  \label{3'} \\
{\tiny :} &&+r^{-1}(+2\eth ^{3}\check{\Upsilon}_{3}-6\eth ^{2}\check{\Upsilon%
}_{3}\ \eth ^{3}\check{\Upsilon}_{3}+\eth ^{3}\check{\Upsilon}_{2}\eth ^{2}%
\check{\Upsilon}_{4}+3\eth ^{2}\check{\Upsilon}_{2}\eth ^{3}\check{\Upsilon}%
_{4}+3\eth \check{\Upsilon}_{2}\eth ^{4}\check{\Upsilon}_{4}+\check{\Upsilon}%
_{2}\eth ^{5}\check{\Upsilon}_{4}-2\eth \check{\Upsilon}_{3}\eth ^{4}\check{%
\Upsilon}_{3})  \notag \\
&{\tiny :}&+r^{-2}\eth ^{3}\check{\Upsilon}_{2}=0.  \notag
\end{eqnarray}

These equations have the following relationship to each other: the
application of the operator $r^{2}\partial _{r}\ $\ to the lower of each of
the three consecutive pairs, [using Eqs.(\ref{g0}) -(\ref{g3})], leads to
the equality with the operator $\eth \ $applied to the upper member of the
pair, e.g., $r^{2}\partial _{r}\ $\ applied to Eq.(\ref{3'}) equals $\eth \
\ $applied to Eq.(\ref{2'}).

Furthermore these four evolutionary Bianchi Identities are closely related
to all the other evolution equations by similar identities. For example,
from Eq.(\ref{t2}), i.e., 
\begin{equation}
\widetilde{\delta }U-\Delta \widetilde{\omega }=(\widetilde{\mu }+\gamma -%
\widetilde{\gamma })\widetilde{\omega }-\nu ,
\end{equation}%
we have%
\begin{eqnarray}
&&-\eth \check{\Upsilon}_{3}^{\prime }+\eth \widetilde{\eth }\eth \check{%
\Upsilon}_{4}+2\check{\Upsilon}_{3}\eth ^{3}\check{\Upsilon}_{4}-2\eth 
\check{\Upsilon}_{4}\eth ^{2}\check{\Upsilon}_{3}  \label{t2'} \\
&{\tiny :}&+r^{-1}(2\eth \check{\Upsilon}_{3}-2\eth \check{\Upsilon}_{3}\eth
^{2}\check{\Upsilon}_{3}+\check{\Upsilon}_{2}\eth ^{3}\check{\Upsilon}%
_{4}+\eth \check{\Upsilon}_{2}\eth ^{2}\check{\Upsilon}_{4})+r^{-2}\eth 
\check{\Upsilon}_{2}=0.  \notag
\end{eqnarray}

The application of $\eth \ $to (\ref{t2'}) yields Eq.(\ref{ts4}),

\begin{equation}
\widetilde{\delta }\widetilde{\nu }-\Delta \widetilde{\mu }=\widetilde{\mu }%
^{2}+\widetilde{\mu }(\gamma +\overline{\gamma })-2\widetilde{\nu }%
\widetilde{\beta }\ ,
\end{equation}%
$\ \ \ \ \ \ \ \ \ \ \ \ $

\noindent or%
\begin{eqnarray}
&&  \label{ts4'} \\
&&-\eth ^{2}\check{\Upsilon}_{3}^{\prime }\ +\eth ^{2}\widetilde{\eth }\eth 
\check{\Upsilon}_{4}+2\check{\Upsilon}_{3}\eth ^{4}\check{\Upsilon}%
_{4}+2\eth ^{2}\check{\Upsilon}_{4}-2\eth ^{2}\check{\Upsilon}_{4}\eth ^{2}%
\check{\Upsilon}_{3}-2\eth \check{\Upsilon}_{4}\eth ^{3}\check{\Upsilon}%
_{3}+2\eth \check{\Upsilon}_{3}\eth ^{3}\check{\Upsilon}_{4}  \notag \\
&&+r^{-1}(2\eth ^{2}\check{\Upsilon}_{3}-2\eth \check{\Upsilon}_{3}\eth ^{3}%
\check{\Upsilon}_{3}-2\eth ^{2}\check{\Upsilon}_{3}\eth ^{2}\check{\Upsilon}%
_{3}+2\eth \check{\Upsilon}_{2}\eth ^{3}\check{\Upsilon}_{4}+\check{\Upsilon}%
_{2}\eth ^{4}\check{\Upsilon}_{4}\ +\eth ^{2}\check{\Upsilon}_{2}\eth ^{2}%
\check{\Upsilon}_{4})  \notag \\
&{\tiny :}&=-r^{-2}\eth ^{2}\check{\Upsilon}_{2}.  \notag
\end{eqnarray}

Finally, in turn, the application of $\eth \ $to (\ref{ts4'}) yields the
Bianchi Identity (\ref{3'}). \ The commutator on a spin-weight$-s$ function $%
W_{s},\ $i.e.,$\ $%
\begin{equation}
\eth \widetilde{\eth }W_{s}-\widetilde{\eth }\eth W_{s}=-2sW_{s}
\end{equation}%
\ has been used several times.

Returning to the evolutionary Bianchi Identities - leaving the first one, (%
\ref{0'}), unchanged - we can see that the last three are simplified by
performing the angular integrations. They can each be rewritten as

\begin{eqnarray}
\eth B_{1} &=&0,  \label{PB1} \\
\eth ^{2}B_{2} &=&0,  \label{PB2} \\
\eth ^{3}B_{3} &=&0,  \label{PB3}
\end{eqnarray}%
with%
\begin{eqnarray}
&&  \label{B1} \\
B_{1} &\equiv &-\check{\Upsilon}_{1}^{\prime }+\widetilde{\eth }\eth \check{%
\Upsilon}_{2}-2\check{\Upsilon}_{2}-2\eth \check{\Upsilon}_{4}\eth \check{%
\Upsilon}_{1}+2\check{\Upsilon}_{3}\eth ^{2}\check{\Upsilon}_{2}-2\eth 
\check{\Upsilon}_{3}\eth \check{\Upsilon}_{2}+2\eth ^{2}\check{\Upsilon}_{3}%
\check{\Upsilon}_{2}  \notag \\
&&+r^{-1}(\check{\Upsilon}_{2}\eth ^{2}\check{\Upsilon}_{2}-2\check{\Upsilon}%
_{1}-2\eth \check{\Upsilon}_{2}\eth \check{\Upsilon}_{2}\ +2\check{\Upsilon}%
_{1}\eth ^{2}\check{\Upsilon}_{3}-2\eth \check{\Upsilon}_{1}\eth \check{%
\Upsilon}_{3}+\check{\Upsilon}_{0}\eth ^{2}\check{\Upsilon}%
_{4})+r^{-2}\Upsilon _{0},  \notag
\end{eqnarray}

\begin{eqnarray}
&&  \label{B2} \\
B_{2} &\equiv &-\check{\Upsilon}_{2}^{\prime }+\widetilde{\eth }\eth \check{%
\Upsilon}_{3}-2\check{\Upsilon}_{3}+\check{\Upsilon}_{2}\eth ^{2}\check{%
\Upsilon}_{4}-2\eth \check{\Upsilon}_{4}\eth \check{\Upsilon}_{2}-\eth 
\check{\Upsilon}_{3}\eth \check{\Upsilon}_{3}+2\check{\Upsilon}_{3}\eth ^{2}%
\check{\Upsilon}_{3}  \notag \\
&&+r^{-1}(\check{\Upsilon}_{1}\eth ^{2}\check{\Upsilon}_{4}-2\eth \check{%
\Upsilon}_{3}\eth \check{\Upsilon}_{2}+\check{\Upsilon}_{2}\eth ^{2}\check{%
\Upsilon}_{3})+r^{-2}\check{\Upsilon}_{1},  \notag
\end{eqnarray}

\begin{equation}
B_{3}\equiv -\check{\Upsilon}_{3}^{\prime }+\widetilde{\eth }\eth \check{%
\Upsilon}_{4}+2\check{\Upsilon}_{3}\eth ^{2}\check{\Upsilon}_{4}-2\eth 
\check{\Upsilon}_{4}\eth \check{\Upsilon}_{3}+r^{-1}(2\check{\Upsilon}%
_{3}-\eth \check{\Upsilon}_{3}\eth \check{\Upsilon}_{3}+\check{\Upsilon}%
_{2}\eth ^{2}\check{\Upsilon}{}_{4})+r^{-2}\check{\Upsilon}_{2},  \label{B3}
\end{equation}

\noindent so that they immediately integrate to%
\begin{eqnarray}
B_{1} &=&K_{1}, \\
B_{2} &=&K_{2}, \\
B_{3} &=&K_{3},
\end{eqnarray}%
with ($K_{1},K_{2},K_{3}$), the kernels of the operators ($\eth ,\eth
^{2},\eth ^{3}$), i.e., ($\eth K_{1},\eth ^{2}K_{2},\eth ^{3}K_{3}$) $=0.\ \ 
$

The kernels, using Eq.(\ref{g.ladder}), are given by

\begin{eqnarray}
K_{3} &=&A(\widetilde{\zeta })+B(\widetilde{\zeta })P+C(\widetilde{\zeta }%
)P^{2}+r^{-1}(D(\widetilde{\zeta })+E(\widetilde{\zeta })P)+r^{-2}F(%
\widetilde{\zeta }), \\
K_{2} &=&-D(\widetilde{\zeta })-E(\widetilde{\zeta })P-2r^{-1}F(\widetilde{%
\zeta }), \\
K_{1} &=&2F(\widetilde{\zeta }).
\end{eqnarray}%
with six arbitrary functions of $(\widetilde{\zeta }).\ $All have angular
(or wire) singularities. Since we wish to restrict ourselves to regular
solutions, the $K_{i}$ are set to zero here.

Our final evolution equations for the potentials then are $%
B_{3}=B_{2}=B_{1}=B_{0}=0\ $or%
\begin{eqnarray}
&&  \label{00'} \\
&&\check{\Upsilon}_{0}^{\prime }-\widetilde{\eth }\eth \check{\Upsilon}_{1}-2%
\check{\Upsilon}_{3}\eth ^{2}\check{\Upsilon}_{1}+2\eth \check{\Upsilon}%
_{3}\eth \check{\Upsilon}_{1}-3\check{\Upsilon}_{2}\eth ^{2}\check{\Upsilon}%
_{2}+2\eth \check{\Upsilon}_{4}\eth \check{\Upsilon}_{0}+\eth ^{2}\check{%
\Upsilon}_{4}\check{\Upsilon}_{0}  \notag \\
&&-r^{-1}(3\check{\Upsilon}_{0}\eth ^{2}\check{\Upsilon}_{3}-2\eth \check{%
\Upsilon}_{3}\eth \check{\Upsilon}_{0}-4\check{\Upsilon}_{0}+\check{\Upsilon}%
_{2}\eth ^{2}\check{\Upsilon}_{1}-6\eth \check{\Upsilon}_{2}\eth \check{%
\Upsilon}_{1}+3\eth ^{2}\check{\Upsilon}_{2}\check{\Upsilon}_{1})  \notag \\
&{\tiny :}&-\frac{\partial }{\partial r}\check{\Upsilon}_{0}-r\eth ^{2}%
\check{\Upsilon}_{4}\frac{\partial }{\partial r}\check{\Upsilon}_{0}=0, 
\notag
\end{eqnarray}

\begin{eqnarray}
&&  \label{11'} \\
&&-\check{\Upsilon}_{1}^{\prime }+\widetilde{\eth }\eth \check{\Upsilon}%
_{2}-2\check{\Upsilon}_{2}-2\eth \check{\Upsilon}_{4}\eth \check{\Upsilon}%
_{1}+2\check{\Upsilon}_{3}\eth ^{2}\check{\Upsilon}_{2}-2\eth \check{\Upsilon%
}_{3}\eth \check{\Upsilon}_{2}+2\eth ^{2}\check{\Upsilon}_{3}\check{\Upsilon}%
_{2}  \notag \\
&&+r^{-1}(-2\check{\Upsilon}_{1}+\check{\Upsilon}_{2}\eth ^{2}\check{\Upsilon%
}_{2}-2\eth \check{\Upsilon}_{2}\eth \check{\Upsilon}_{2}\ +2\check{\Upsilon}%
_{1}\eth ^{2}\check{\Upsilon}_{3}-2\eth \check{\Upsilon}_{1}\eth \check{%
\Upsilon}_{3}+\check{\Upsilon}_{0}\eth ^{2}\check{\Upsilon}%
_{4})+r^{-2}\Upsilon _{0}=0,  \notag
\end{eqnarray}

\begin{eqnarray}
&&-\check{\Upsilon}_{2}^{\prime }+\widetilde{\eth }\eth \check{\Upsilon}%
_{3}-2\check{\Upsilon}_{3}+\check{\Upsilon}_{2}\eth ^{2}\check{\Upsilon}%
_{4}-2\eth \check{\Upsilon}_{4}\eth \check{\Upsilon}_{2}-\eth \check{\Upsilon%
}_{3}\eth \check{\Upsilon}_{3}+2\check{\Upsilon}_{3}\eth ^{2}\check{\Upsilon}%
_{3}  \label{22'} \\
&&+r^{-1}(\check{\Upsilon}_{1}\eth ^{2}\check{\Upsilon}_{4}-2\eth \check{%
\Upsilon}_{3}\eth \check{\Upsilon}_{2}+\check{\Upsilon}_{2}\eth ^{2}\check{%
\Upsilon}_{3})+r^{-2}\check{\Upsilon}_{1}=0,  \notag
\end{eqnarray}

\begin{eqnarray}
&&  \label{33'} \\
-\check{\Upsilon}_{3}^{\prime }+\widetilde{\eth }\eth \check{\Upsilon}_{4}+2%
\check{\Upsilon}_{3}\eth ^{2}\check{\Upsilon}_{4}-2\eth \check{\Upsilon}%
_{4}\eth \check{\Upsilon}_{3}+r^{-1}(2\check{\Upsilon}_{3}-\eth \check{%
\Upsilon}_{3}\eth \check{\Upsilon}_{3}+\check{\Upsilon}_{2}\eth ^{2}\check{%
\Upsilon}{}_{4})+r^{-2}\check{\Upsilon}_{2} &=&0.  \notag
\end{eqnarray}

Finally, using Eqs.(\ref{g2}) and (\ref{g3}) of the ladder relations in Eq.(%
\ref{33'}),\ we obtain our non-linear wave equation $\check{\Upsilon}_{4}\ $%
which carries all the information of the evolution of the individual $\check{%
\Upsilon}_{n};$

\begin{eqnarray}
&&\partial _{r}\check{\Upsilon}_{4}^{\prime }-r^{-2}\widetilde{\eth }\eth 
\check{\Upsilon}_{4}-\partial _{r}^{2}\check{\Upsilon}_{4}-4r^{-1}\partial
_{r}\check{\Upsilon}_{4}  \label{wave eqs} \\
&=&-2\eth \check{\Upsilon}_{4}\ \partial _{r}\eth \check{\Upsilon}%
_{4}+4\partial _{r}\check{\Upsilon}_{4}\eth ^{2}\check{\Upsilon}_{4}-r\
\partial _{r}\eth \check{\Upsilon}_{4}\ \partial _{r}\eth \check{\Upsilon}%
_{4}+r\ \partial _{r}^{2}\check{\Upsilon}_{4}\ \eth ^{2}\check{\Upsilon}_{4}.
\notag
\end{eqnarray}

Since, with the ladder relations, Eq.(\ref{wave eqs}) is equivalent to the
previous four equations, one could use either for integration.

\section{Examples}

A small set of solutions can be found by make the starting ansatz

\begin{equation}
\check{\Upsilon}_{4}=c_{4}-c_{3}r^{-1}+\frac{1}{2}c_{2}r^{-2}-\frac{1}{6}%
c_{1}r^{-3}+\frac{1}{24}c_{0}r^{-4}.  \label{gamma4}
\end{equation}%
This leads immediately to

\begin{eqnarray}
\check{\Upsilon}_{0} &=&c_{0},  \label{c's} \\
\check{\Upsilon}_{1} &=&c_{1}-c_{0}r^{-1},  \notag \\
\check{\Upsilon}_{2} &=&c_{2}-c_{1}r^{-1}+\frac{1}{2}c_{0}r^{-2},  \notag \\
\check{\Upsilon}_{3} &=&c_{3}-c_{2}r^{-1}+\frac{1}{2}c_{1}r^{-2}-\frac{1}{6}%
c_{0}r^{-3},  \notag
\end{eqnarray}%
the $c\ $'s being independent of $r.\ \ $Inserting $\check{\Upsilon}_{4}\ $%
into (\ref{33'}), and equating powers of $r^{-1},\ $yields the evolution
equations for the individual $c\ $'s:%
\begin{eqnarray}
c_{0}^{\prime }-\widetilde{\eth }\eth c_{1}-2c_{3}\eth ^{2}c_{1}+2\eth
c_{3}\eth c_{1}-3c_{2}\eth ^{2}c_{2}+2\eth c_{4}\eth c_{0}+c_{0}\eth
^{2}c_{4} &=&0,  \label{c' s} \\
-c_{1}^{\prime }+\widetilde{\eth }\eth c_{2}-2c_{2}-2\eth c_{4}\eth
c_{1}+2c_{3}\eth ^{2}c_{2}-2\eth c_{3}\eth c_{2}+2c_{2}\eth ^{2}c_{3} &=&0, 
\notag \\
-c_{2}^{\prime }+\widetilde{\eth }\eth c_{3}-2c_{3}+c_{2}\eth
^{2}c_{4}-2\eth c_{4}\eth c_{2}-\eth c_{3}\eth c_{3}+2c_{3}\eth ^{2}c_{3}
&=&0,  \notag \\
-c_{3}^{\prime }+\widetilde{\eth }\eth c_{4}+2c_{3}\eth ^{2}c_{4}-2\eth
c_{4}\eth c_{3} &=&0.  \notag
\end{eqnarray}

We obtain a system easily integrated if we make the further ansatz, 
\begin{equation}
c_{4}=0.  \label{c4=0}
\end{equation}

The (\ref{c' s}) reduce to

\begin{eqnarray}
c_{0}^{\prime } &=&\widetilde{\eth }\eth c_{1}+2c_{3}\eth ^{2}c_{1}-2\eth
c_{3}\eth c_{1}+3c_{2}\eth ^{2}c_{2},  \label{c'=} \\
c_{1}^{\prime } &=&\widetilde{\eth }\eth c_{2}-2c_{2}+2c_{3}\eth
^{2}c_{2}-2\eth c_{3}\eth c_{2}+2c_{2}\eth ^{2}c_{3},  \notag \\
c_{2}^{\prime } &=&\widetilde{\eth }\eth c_{3}-2c_{3}-\eth c_{3}\eth
c_{3}+2c_{3}\eth ^{2}c_{3},  \notag \\
c_{3}^{\prime } &=&0.  \notag
\end{eqnarray}%
leading to%
\begin{eqnarray}
c_{3} &=&c_{3}^{0}(\zeta ,\widetilde{\zeta }),  \label{a solution} \\
c_{2} &=&c_{2}^{0}(\zeta ,\widetilde{\zeta })+u(\widetilde{\eth }\eth
c_{3}^{0}-2c_{3}^{0}-\eth c_{3}^{0}\eth c_{3}^{0}+2c_{3}^{0}\eth
^{2}c_{3}^{0})  \notag \\
c_{1} &=&c_{1}^{0}(\zeta ,\widetilde{\zeta })+\int^{u}du(\widetilde{\eth }%
\eth c_{2}-2c_{2}+2c_{3}^{0}\eth ^{2}c_{2}-2\eth c_{3}^{0}\eth
c_{2}+2c_{2}\eth ^{2}c_{3}^{0}),  \notag \\
c_{0} &=&c_{0}^{0}(\zeta ,\widetilde{\zeta })+\int^{u}du(\widetilde{\eth }%
\eth c_{1}+2c_{3}^{0}\eth ^{2}c_{1}-2\eth c_{3}^{0}\eth c_{1}+3c_{2}\eth
^{2}c_{2}).  \notag
\end{eqnarray}%
\qquad

These solutions have in general cubic $u\ $dependence.\ 

The special case of $c_{3}=c_{2}=c_{1}=0,\ $leaves $c_{0}=c_{0}^{0}(\zeta ,%
\widetilde{\zeta }),\ $a time independent solution

\begin{eqnarray}
\check{\Upsilon}_{0} &=&c_{0}^{0},  \label{time independent} \\
\check{\Upsilon}_{1} &=&-c_{0}^{0}r^{-1},  \notag \\
\check{\Upsilon}_{2} &=&+\frac{1}{2}c_{0}^{0}r^{-2},  \notag \\
\check{\Upsilon}_{3} &=&-\frac{1}{6}c_{0}^{0}r^{-3}.  \notag
\end{eqnarray}

Another class of solutions of (\ref{c' s}) starts with the ansazt, $%
c_{0}=c_{1}=c_{2}=c_{3}=0.\ $The last of (\ref{c' s}), becomes%
\begin{equation}
\widetilde{\eth }\eth c_{4}=\partial _{\widetilde{\zeta }}(P^{4}\partial
_{\zeta }(P^{-2}c_{4}))=0.  \label{typeN}
\end{equation}

\bigskip The solutions, all type N, given by 
\begin{equation}
c_{4}=P^{2}G(\widetilde{\zeta })+P^{2}\int P^{-4}\partial _{\zeta }F(\zeta
)d\zeta ,
\end{equation}%
are all singular, and have, unfortunately, wire singularities.

\section{Discussion}

In the interests of honesty, we describe the thought perturbations that led
to the present investigation. Several months ago we were working on aspects
of the complexified full Einstein equations - in particular - on asymptotic
shear-free null geodesic congruences, and simply noticed that we could, with
great ease, integrate the radial Bianchi Identities via the introduction of
the five potentials (described in the text) with their ladder relations.
Though we initially had no interest in investigating the self-dual Einstein
metrics, the apparent simplicity, coming from the use of the potentials,
charmed us into going for the "entire thing". Briefly it even seemed
possible that we might go the entire way and be able to construct \textit{%
explicit} vacuum self-dual metrics directly from the initial data. It turned
out that this hope was very much dashed: first of all by the complexity of
the remaining equations - they involved long and hard calculations - and
more important, we ended with the unpleasant non-linear wave equation not
easily solvable.

So in the end the question remains: though we have found some pretty
results, what did we really accomplish. These results certainly clarify the
structure of the self-dual equations - but do the results have applications
or are they of interest in other investigations? \ They appear to
potentially have use for our own investigations - but we are not certain.

There are several further issues to be mentioned.

The first is a mild mystery. \ The data needed for solutions to our
non-linear wave equation are \textit{two complex functions}, one the value
of $\check{\Upsilon}_{4}\ $on an initial null cone $u_{0},\ $i.e.,\ an
arbitrary function of ($r,\zeta ,\widetilde{\zeta }$)\ and the second, $%
c_{4},\ $(the\ asymptotic value of $\check{\Upsilon}_{4}$) a news-like
function of ($u,\zeta ,\widetilde{\zeta }$)\ that drives the evolution. \ On
the other hand, from the original formulation of the self-dual Einstein
metrics, via the so-called `good-cut' equation, one needs only one complex
function as the needed data, namely the asymptotic Bondi shear $\sigma
^{0}(u_{B},\zeta ,\widetilde{\zeta }),\ $with $u_{B}\ $the Bondi time$.\ $\
It is probable that by the proper counting of the ranges of the arguments it
can be shown that the two sets contain the same amount of data.

A second is that there should be a geometric meaning to the potentials, $%
\check{\Upsilon}_{n}$. We have not attempted to investigate that.

For whatever it might be worth, and it is suggestive that it is worth
something, we point out that studies$^{18}$ of the self-dual Yang-Mills
equations in the spin-coefficient form show that, for any gauge group, the
entire set of Yang-Mills equations can be encapsulated into a single
nonlinear wave equation very similar to our non-linear equation, (\ref{wave
eqs}). \ The basic variable, $\mathfrak{F}_{2}$, is one of three \textit{%
matrix} valued\ potentials, $\mathfrak{F}_{i}$, all connected to each other
by a ladder relationship analogous to the one for GR.

The Yang-Mills wave equation$^{19}$ is%
\begin{equation}
\partial _{r}\mathfrak{F}_{2}^{\prime }-\partial _{r}^{2}\mathfrak{F}_{2}\ 
\mathfrak{-}2r^{-1}\partial _{r}\mathfrak{F}_{2}\ \mathfrak{-}r^{-2}%
\widetilde{\eth }\eth \mathfrak{F}_{2}=[\eth \mathfrak{F}_{2}\mathfrak{,}%
\partial _{r}\mathfrak{F}_{2}],  \label{YM}
\end{equation}%
with ladder relations among the potentials%
\begin{eqnarray}
\mathfrak{F}_{1} &=&r^{2}\partial _{r}\mathfrak{F}_{2}, \\
\mathfrak{F}_{0} &=&r^{2}\partial _{r}\mathfrak{F}_{1}=r^{2}\partial
_{r}(r^{2}\partial _{r}\mathfrak{F}_{2}).
\end{eqnarray}

\noindent\ \qquad The tetrad components of the self-dual Yang-Mills field
are given by%
\begin{eqnarray}
\widetilde{\chi }_{0} &=&r^{-3}\mathfrak{F}_{0}, \\
\widetilde{\chi }_{1} &=&r^{-2}\eth \mathfrak{F}_{1}, \\
\widetilde{\chi }_{2} &=&r^{-1}\eth ^{2}\mathfrak{F}_{2},
\end{eqnarray}%
in an almost perfect analogy with the self-dual gravitational case.

\end{document}